\documentclass[letter]{article}
\usepackage{spconf,graphicx,amssymb,amsmath,booktabs,hyperref,todonotes,scalerel,stackengine}

\newcommand{\beq}{\begin{equation}}
\newcommand{\eeq}{\end{equation}}
\newcommand{\tr}[1]{{#1}^\top}
\newcommand{\vect}[1]{\mathbf{#1}}
\newcommand{\mr}[1]{\mathrm{#1}}

\newcommand{\Real}{\mathbb{R}}
\newcommand{\rref}{{(0)}}

\stackMath
\newcommand\reallywidehat[1]{%
\savestack{\tmpbox}{\stretchto{%
  \scaleto{%
    \scalerel*[\widthof{\ensuremath{#1}}]{\kern-.6pt\bigwedge\kern-.6pt}%
    {\rule[-\textheight/2]{1ex}{\textheight}}
  }{\textheight}%
}{0.5ex}}%
\stackon[1pt]{#1}{\tmpbox}%
}

\newcommand{\GG}{\mathcal{G}}
\newcommand{\VV}{\mathcal{V}}
\newcommand{\EE}{\mathcal{E}}
\newcommand{\NN}{\mathcal{N}}
\newcommand{\LLambda}{\mathbf{\Lambda}}
\newcommand{\TTheta}{\boldsymbol{\theta}}

\newcommand{\LL}{\vect{L}}
\newcommand{\II}{\vect{I}}

\newcommand{\uhat}{\reallywidehat{\vect{u}}}
\newcommand{\xx}{\vect{x}}

\newcommand{\UU}{\vect{U}}
\newcommand{\TT}{\vect{T}}

\newcommand{\XX}{\vect{X}}

\newcommand{\Uhat}{{\reallywidehat{\vect{U}}}}
\newcommand{\Utilde}{\widetilde{\vect{U}}}

\newcommand{\DD}{\vect{D}}

\newcommand{\WW}{\vect{W}}
\newcommand{\AAA}{\vect{A}}
\newcommand{\mmu}{\boldsymbol{\mu}}

\DeclareMathOperator*{\argmin}{arg\,min}

\pdfoutput=1

\title{Spectral Graph Transformer Networks for~Brain~Surface~Parcellation}

\name{ 
      Ran He$^{\dagger\mathparagraph}$\sthanks{Equal contribution of R. He and K. Gopinath.}\qquad
      Karthik Gopinath$^{\dagger\ast}$ \qquad
      Christian Desrosiers$^{\dagger}$ \qquad 
      Herve Lombaert$^{\dagger}$}

\address{$^{\dagger}$ ETS Montr\'eal, Canada \\
    $^{\mathparagraph}$ Beijing Institute of Technology, China}

\begin{document}
\maketitle         

\begin{abstract}

The analysis of the brain surface modeled as a graph mesh is a challenging task. Conventional deep learning approaches often rely on data lying in the Euclidean space. As an extension to irregular graphs, convolution operations are defined in the Fourier or spectral domain. This spectral domain is obtained by decomposing the graph Laplacian, which captures relevant shape information. However, the spectral decomposition across different brain graphs causes inconsistencies between the eigenvectors of individual spectral domains, causing the graph learning algorithm to fail. Current spectral graph convolution methods handle this variance by separately aligning the eigenvectors to a reference brain in a slow iterative step. This paper presents a novel approach for learning the transformation matrix required for aligning brain meshes using a direct data-driven approach. Our alignment and graph processing method provides a fast analysis of brain surfaces. The novel Spectral Graph Transformer (SGT) network proposed in this paper uses very few randomly sub-sampled nodes in the spectral domain to learn the alignment matrix for multiple brain surfaces. We validate the use of this SGT network along with a graph convolution network to perform cortical parcellation. Our method on 101 manually-labeled brain surfaces shows improved parcellation performance over a no-alignment strategy, gaining a significant speed (1400 fold) over traditional iterative alignment approaches.
\end{abstract}

\begin{keywords}
Spectral transformer network, Cortical parcellation, Graph Convolution Network
\end{keywords}

\section{Introduction}

The surface of a human brain is a complex geometrical structure containing multiple convoluted folding patterns. Statistical analysis of the brain surface aids in understanding its anatomy, and machine learning methods are often sought for automating this analysis. Conventional machine learning frameworks exploit spatial information from the Euclidean domain such as image or volumetric coordinates \cite{Zhang2011ODVBA,Hua2013Unbiased}. Similarly, state-of-the-art deep learning approaches \cite{dolz20173d,Kamnitsas2017Efficient} operate on data lying in Euclidean spaces, offering a drastic speed advantage over traditional methods. However, the geometry of the brain is highly variable, hindering the direct use of these modern deep learning algorithms over multiple brain surfaces. 

Recently, deep learning approaches on irregular graphs \cite{Bronstein2017Geometric,Monti2017Geometric,Levie2018CayleyNets} have been proposed. These methods formulate a convolution theorem from Fourier space to spectral domains over graphs. One main limitation of these spectral approaches is their lack of expressing surface data in comparable spectral bases across different surface domains \cite{Bronstein2013Making,Kovnatsky2013Coupled,Eynard2015Multimodal}. The Laplacian eigenbases are indeed incompatible across multiple geometries, challenging their direct use during training. As a solution, some recent work \cite{Masci2015Geodesic,Boscaini2016Learning} maps the local information onto geodesic patches and uses conventional template matching in spatial convolutions. For instance, \cite{Monti2017Geometric} proposed local convolution operation as filtering over small neighborhoods in spatial domain. Their spatial representations of surface data remain, however, defined in a Euclidean space by using polar representations of pixels or mesh vertices. 

\begin{figure*}[th]
 \centering
  
  
    \includegraphics[width=0.9\textwidth]{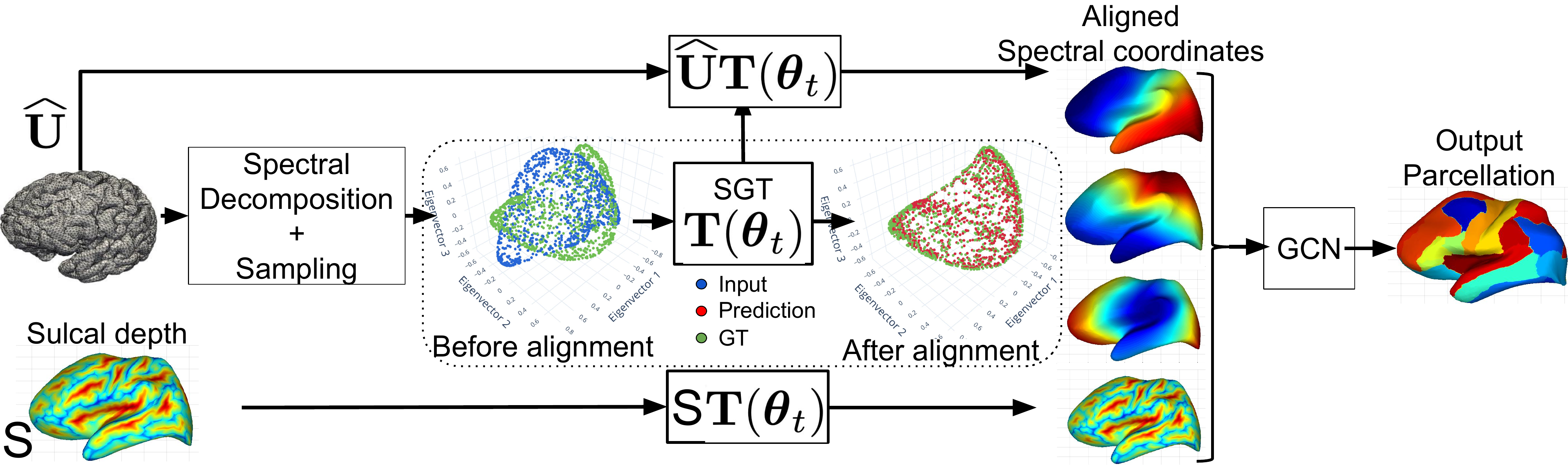}
    
  \caption{\textbf{Overview of the our architecture:} The spectral decomposition of the brain graph is randomly sub-sampled as an input point cloud to a SGT network. The SGT learns the transformation parameters aligning the eigenvectors of multiple brains. The transformation matrix is multiplied with original spectral coordinates to feed the GCN for parcellation. The point cloud is illustrated before and after alignment with our SGT network. The GCN architecture follows recommendations from \cite{gopinath2019graph}.}
  \label{spe_tra_arch}
\end{figure*}

In the literature, spectral graph matching has been used to transfer surface data across aligned spectral domains \cite{Lombaert2015Brain}. Such strategy \cite{gopinath2019graph} enables the learning of spectral graph convolution networks across multiple surface data. These methods, however, involve an explicit computation of a transformation map for each brain towards one reference template. This process of aligning the eigenvectors of graph Laplacians is currently an important computational bottleneck. This expensive step is necessary in such approach to handle the differences across eigenvectors, including sign flips, ordering, and mixing of eigenvectors in higher frequencies. In this work, we propose a framework for learning this transformation function across multiple brain surfaces. In an alternative application for natural image classification, \cite{jaderberg2015spatial} proposes a transformer network for CNNs for learning a transformation matrix to spatially standardize the image data. Similarly, \cite{qi2017pointnet} also proposes a transformation network for learning over point clouds of geometric structures. These methods are, however, limited to pointwise information in a Euclidean space. This paper introduces a Spectral Graph Transformer Network (SGT) to learn the parameters for aligning multiple surfaces directly in the spectral domain. We illustrate the learning capabilities of this approach with an application to brain parcellation. We use the aligned coordinates from our SGT network along with a graph convolution network (GCN) for quantifying parcellation. The learnt alignment of 101 manually-labeled brain surfaces \cite{Klein2017Mindboggling} reveals that our approach improves brain parcellation by 4.4$\%$, from an average Dice overlap of 78.8$\%$ to 83.2$\%$. The performance of our method is shown to be at par with traditional alignment strategies, performing at 84.4$\%$, but gains a significant speed improvement. The learning of an end-to-end SGT and GCN model enables a direct, automatic learning of surface data across multiple brains. Our SGT part learns a transformation matrix that handles the eigenvector differences, while the GCN part focuses on the brain parcellation. The next section details the fundamentals of our SGT and GCN model, followed by an evaluation of our alignment strategy for graph convolutions.

\section{Method}

An overview of the method is shown in Fig.~\ref{spe_tra_arch}. Firstly, the cortical surfaces modeled as brain graphs are embedded in a spectral manifold using the graph Laplacian operator. Secondly, graph nodes are randomly sampled in the spectral embeddings and fed to the SGT network to align the brain embeddings. Finally, a GCN provides a labeled graph as output, taking spectral coordinates and cortical sulcal depth as input. 

\subsection{Spectral embedding of brain graphs}

Let $\GG = \{\VV, \EE\}$ be a brain graph defined with node set $\VV$, such that $|\VV|=N$, and edge set $\EE$. Each node $i$ has a feature vector $\xx_i \in \Real^3$ representing its 3D coordinates. We map $\GG$ to a low-dimension manifold using the normalized graph Laplacian operator $\LL = \II - \DD^{-\frac{1}{2}}\AAA\DD^{-\frac{1}{2}}$, where $\AAA$ is the weighted adjacency matrix and $\DD$ the diagonal degree matrix. In this work, we define the weight between two adjacent nodes as the inverse of their Euclidean distance. Let $\LL = \UU \LLambda \tr{\UU}$ be the eigendecomposition of $\LL$, the normalized spectral coordinates of nodes are given by $\Uhat = \LLambda^{-\frac{1}{2}} \UU$. 
\subsection{Spectral transformer network}

The normalized spectral coordinates $\Uhat$ from the spectral embedding of $\LL$ is only defined up to an orthogonal transformation. We thus need to align the spectral representations of different brain graphs to a common representation. As a base reference, we align the normalized spectral embedding of all the brain surfaces to a template $\Uhat_\mr{ref}$ in the dataset. This traditional alignment process involves computing an expensive optimal orthogonal transform based on iterative Procustes algorithm \cite{Lombaert2015Brain}, which can be formulated as
\begin{equation}\label{eq:correspondence}
    \argmin_{\pi, \TT} \ \sum_{i=1}^N \big\| \TT \,\uhat_i \,- \,\uhat^\rref_{\pi(i)} \big\|_2^2
\end{equation}
This alignment step is computationally expensive, taking few seconds to converge. Also, the alignment process is independent of the final target task. Our STN consists of learning the transformation matrix $\TT$ for every brain graph in a data-driven manner. As input to the network, we provide $\Uhat_{\mr{sub}}$, a set of $N$ randomly sub-sampled $\Uhat$ coordinates, chosen similarly to \cite{qi2017pointnet}, with enough samples to recover $\TT$. Since most information on graph connectivity is encoded in the first eigencomponents of $\LL$, to limit processing times, we only keep the first $3$ components for the learning step. Thus, $\Uhat_{\mr{sub}}$ is a matrix of size $N\!\times\!3$. Fig.~\ref{spe_tra_arch} describes the architecture of our spatial transformer network. 

The model first applies a sequence of two point-wise linear transformation layers on $\Uhat_{\mr{sub}}$, each one followed by a non-linear rectifier (ReLU) function. Such layer takes a $N\!\times\!M_{l-1}$ matrix $\XX$ as input and post-multiplies it by a $M_{l-1}\!\times\!M_l$ parameter matrix $\WW_l$ to give an output matrix of size $N\!\times\!M_l$. This transformation, which is similar to $1\!\times\!1$ convolutions in CNNs, expresses each embedded node with respect to a shared set of $M_l$ hyper-planes in the spectral space, and is used to capture the global shape of the embedding. In our model, we use $M_1=256$ for the first layer and $M_2=128$ for the second one (note that $M_0=3$). Next, the output of the second point-wise transformation layer is converted to a fixed-size representation of size $128\!\times\!1$ by applying average pooling. Last, to get the final spectral transformation matrix, we apply three MLP layers of size [128, 64, 9], also with ReLU activations, and reshape the output of the last layer into a $3 \times 3$ matrix. This transformation matrix is multiplied to the normalized spectral coordinates $\Uhat$ to obtain the aligned spectral coordinates. 

The parameters of the spectral transformer network are optimized by computing the mean square error between the predicted coordinates and spectral coordinates $\Utilde$ obtained with the iterative alignment method. To enforce regularization during training, and match the possible rotation and flip ambiguity of eigendecomposition, we also add a second loss term imposing the transformation matrix to be orthogonal. The final loss function is given by
\begin{equation}\label{eq:error_sgt}
    E_{\mr{spt}}(\TTheta_{t}) \ =  \|\Utilde -  \Uhat\, \TT(\TTheta_{t})\|^2_F \ + \   \|\TT(\TTheta_{t})\tr{\TT}(\TTheta_{t}) -  \II\|^2_F
\end{equation}


\subsection{Graph convolution on surfaces}


The second part of our end-to-end model is based on a geometric convolutional neural network that maps the now-aligned spectral coordinates to a common comparable graph embedding. A generalized convolution operation on a graph $\GG = \{\VV, \EE\}$, with $\NN_i = \{j \, | \, (i,j) \in \EE\}$, as the neighbors of node $i \in \VV$, is defined as
\begin{equation}\label{eq:convolution}
    z_{ip}^{(l)} \ = \ 
        \sum_{j \in \NN_i} \sum_{q=1}^{M_l} \sum_{k=1}^{K_l} 
            w_{pqk}^{(l)} \cdot y_{jq}^{(l)} \cdot \varphi(\uhat_i, \uhat_j; \, \TTheta^{(l)}_k)  \ + \ b_p^{(l)},
\end{equation}
where $\varphi(\uhat_i, \uhat_j; \TTheta_k)$ is a symmetric kernel in the embedding space with parameter $\TTheta_k$. In this work, we follow \cite{gopinath2019graph} and use a Gaussian kernel: $\varphi(\uhat_i, \uhat_j; \mmu_k, \sigma_k) \ = \ \exp\big(-\sigma_k \, \|(\uhat_j - \uhat_i) - \mmu_k\|^2\big)$.

We define a fully-convolutional network comprising of 4 graph convolution layers with sizes 256, 128, 64, and 32. Each layer have $K_l=6$ Gaussian kernels similar to \cite{gopinath2019graph}. The total target parcels are 32, hence, our last layer is of size 32. Leaky ReLU is applied after each layer to obtain our filter responses. A softmax operation is used after the last graph convolution layer in order to obtain the probabilities of the mutually-exclusive parcels at each node. Our output loss function employs a cross-entropy with Dice loss for all parcellations.
Our final end-to-end model comprising of a spectral transformer and a graph convolution network for brain parcellation is trained using the loss function given by
\begin{equation}\label{eq:cross-entropy}
    E_{\mr{final}}(\TTheta_{t}, \TTheta_{g}) \ = \ \lambda \, E_{\mr{spt}}(\TTheta_{t}) \ + \ E_{\mr{gcn}}(\TTheta_g).
\end{equation}
This final loss $E_{\mr{final}}$ is minimized by back-propagating the error using standard gradient descent optimization.

\begin{figure}[t]
\centering
\includegraphics[width=\linewidth]{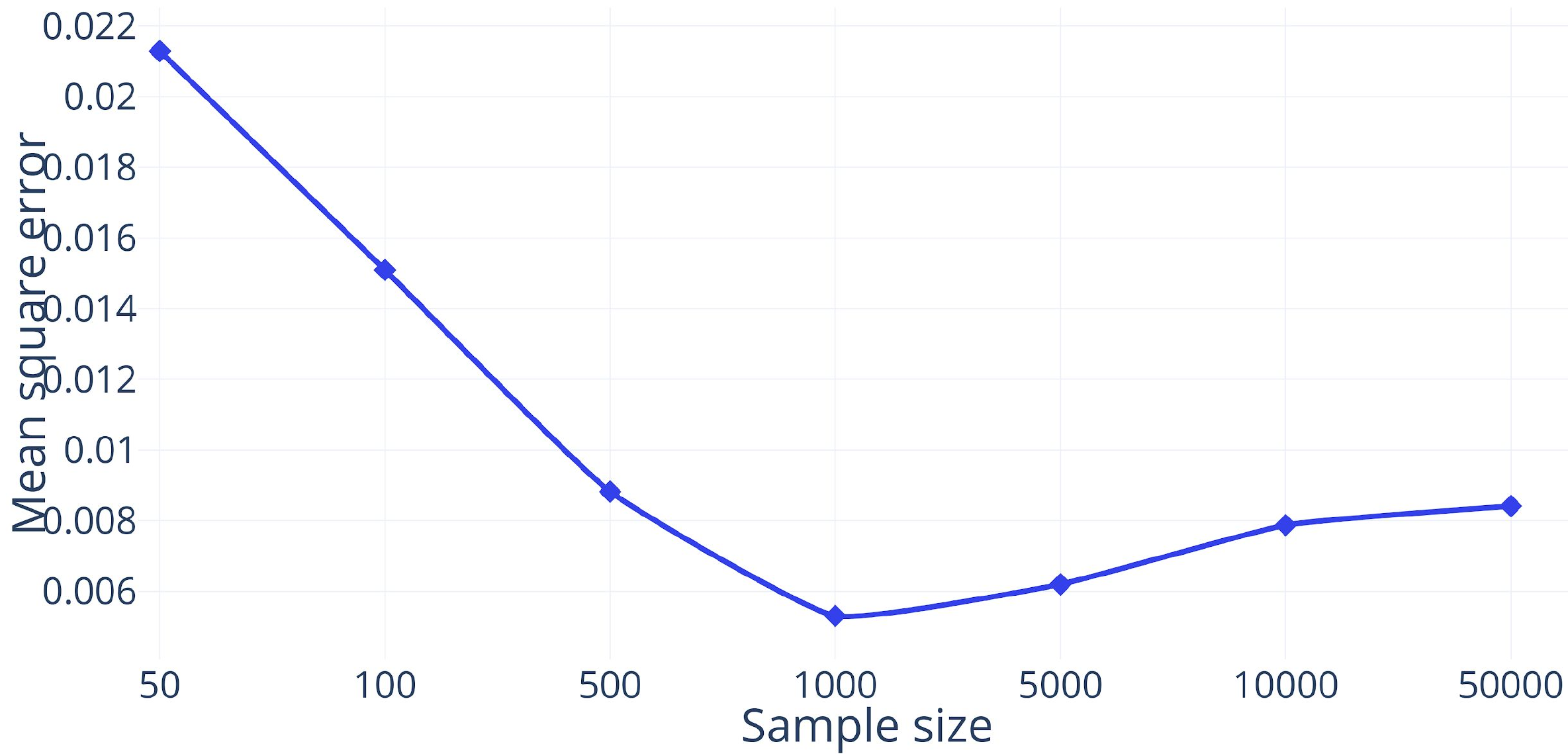}
\caption{\textbf{SGT network data sampling:} Each point indicates the performance of SGT model in terms of mean square error. It is observed that the model trained with fewer nodes than 500 perform poorly compared to all the models. The best mean square error is achieved for model with 1000 nodes as input to SGT.}
\label{plot_sample_size}
\end{figure}

\begin{figure*}
\centering
\includegraphics[width=0.875\linewidth]{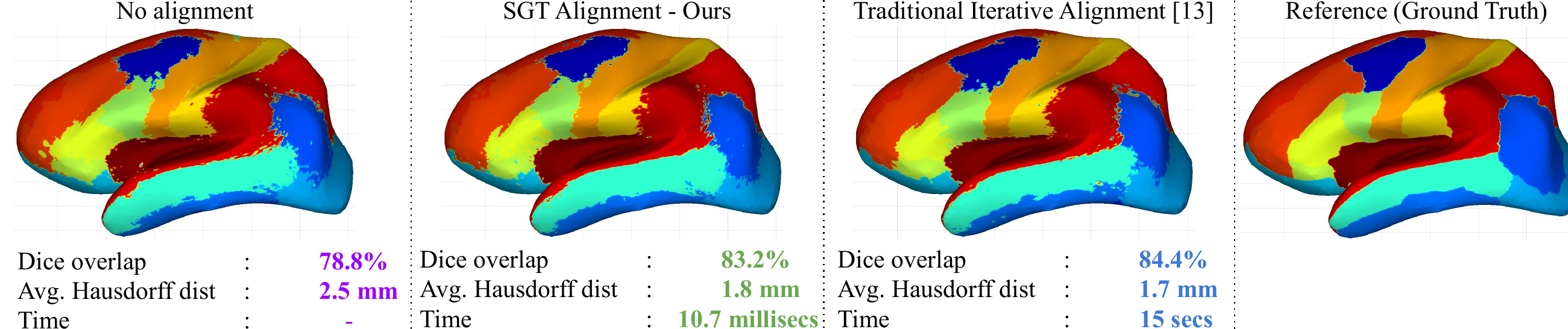}
\caption{\textbf{Brain parcellation:} Performance comparison of different alignment strategies using GCN measured with average Dice overlap and Hausdorff distance. Model trained with no SGT yields low Dice of 78.6$\%$ with irregular segmentation boundaries. Training an end-to-end SGT and GCN model achieves a Dice overlap of 83.2$\%$ similar to the performance of a traditional alignment model with Dice overlap of 84.4$\%$. The Hausdorff distance and qualitative results show very similar results between the two methods. However, a significant speed gain in order of 10.7 \textit{milliseconds} is achieved with our SGT and GCN model.} 
\label{qualitative_parc_results}
\end{figure*}

\begin{table*}[!t]
\centering
\caption{\textbf{Different alignment strategies with GCN approaches} -- Average Dice overlaps (in \%) over 32 parcels on test set are shown along with classification accuracy (in \%), and average Hausdorff distances (in millimeters).}
\label{comp_meth}
\begin{small}
\begin{tabular}{lccc}
\toprule
\textbf{Method}    & \textbf{Dice overlap (\%)}   & \textbf{Accuracy (\%)} &  \textbf{Avg.\,Hausdorff (mm)}\\
\midrule\midrule
No Alignment   & 78.82\,$\pm$\,4.02 & 81.68\,$\pm$\,3.88   & 2.54\,$\pm$\,2.86   \\


Pretrained + Orthogonal    & 81.97\,$\pm$\,3.20 & 84.14\,$\pm$\,2.88 & 1.99\,$\pm$\,2.19   \\

Pretrained + MSE           & 82.29\,$\pm$\,4.46 & 84.38\,$\pm$\,4.09 & 1.94\,$\pm$\,2.23   \\

End-to-end ~ (Ours - 10.7 \textit{milliseconds})  & 83.26\,$\pm$\,3.66 & 85.17\,$\pm$\,3.48 & 1.85\,$\pm$\,2.04   \\

\midrule\midrule

Traditional Alignment \cite{gopinath2019graph} ~ (15 seconds)           & 84.42\,$\pm$\,2.59 & 85.99\,$\pm$\,2.53 & 1.76\,$\pm$\,1.75   \\

\bottomrule
\end{tabular}
\end{small}
\end{table*}

\section{Experiments and Results} 

In this section, we evaluate how inputs affect our SGT network. The optimal SGT parameters are thereafter used to train our end-to-end model for brain parcellation. We validate our approach on the Mindboggle \cite{Klein2017Mindboggling} dataset containing manually-labeled brain surfaces. The dataset contains 101 cortical meshes, each with 102K to 185K vertices and 32 manually-labeled parcels. We randomly split the dataset into training, validation and testing in a 70-10-20$\%$ ratio for our experiments. Here, we induce random sign flips on the eigenvectors of the training dataset to balance flipping and rotation variance. The performance of the methods are measured in terms of average Dice overlap and Hausdorff distances. The experiments are carried out on an i7 desktop computer with 16GB of RAM and a Nvidia Titan Xp GPU.

\subsection{Spectral transform data sampling}

Our spectral transformer network takes as input a set of points in the spectral domain. The number of eigenvectors is fixed to three, as suggested in \cite{gopinath2019graph}. To evaluate the effect of input size $N$, we sample spectral points randomly from $50$ to $50,000$. We study the performance of spectral alignment using our SGT model in terms of mean square error.

The results shown in Fig.~\ref{plot_sample_size} illustrate that the best alignment performance is achieved with a sub-sampling size of $N = 1000$. The input data with $N = 50, 100, 500$ is inadequate to capture the complete geometric information of the brain, as seen in Fig.~\ref{plot_sample_size}. In addition to lower performance, a higher number of nodes also increases memory consumption and computation time. The gain in mean square error for input size over $N = 1000$ can also be seen in Fig.~\ref{plot_sample_size}.

\subsection{Brain surface parcellation}

We now evaluate the performance of our end-to-end SGT and GCN model on brain surface parcellation. The predicted transformation matrix from SGT aligns all brain surfaces. The number of embedded node coordinates used during training SGT is set to $N = 1000$. These nodes are randomly sub-sampled for each subject during the training of our end-to-end model.

Our method is compared with different alignment strategies for graph parcellation. We show the limitations of ignoring the spectral alignment. The GCN trained with non-aligned spectral coordinates achieves a Dice overlap performance of 78.8$\%$. This low accuracy is due to the incompatibility of eigenbases across brain surfaces. Training our end-to-end SGT with GCN provides a performance improvement of 4.4$\%$ for parcellation over no alignment. 
Next, our transformer network is trained independently from the parcellation task in order to learn the SGT weights. The rationale of this experiment is to evaluate the use of a fixed alignment strategy for learning the GCN model. We evaluate the use of both SGT loss and orthogonal regularization independently. The model trained with only orthogonal regularization has a performance gain of 3.1$\%$ from 78.8$\%$ to 81.9$\%$. This increase indicates the usefulness of regularization to learn rotation and flipping. We see a further performance boost by training our SGT model with mean square error. Table \ref{comp_meth} shows a similar performance gain of 3.4$\%$ compared to not using alignment. Note that updating the weights of both SGT and GCN in an end-to-end framework further guides the learning of the transformation matrix. This experiment setup trains the SGT model to learn a transformation most suitable for the parcellation task. Our end-to-end model indeed yields an improvement in average Dice overlap of 83.4$\%$ compared to 82.2$\%$ when trained separately. The results of the experiments are reported in Table~\ref{comp_meth}.

\section{Conclusion} 
\label{sec:diss_con}

This paper presented a novel end-to-end framework for learning a spectral transformation required for graph convolution networks. The proposed SGT network learns a transformation in the spectral domain that maps input spectral coordinates to a reference set. We first evaluate the optimal size of the coordinate set necessary for training the SGT network. Next, our experiments on brain surface parcellation validate the benefits of our alignment strategy. Training a GCN model without any alignment results in a low Dice overlap and irregular parcel boundaries as shown in Fig.~\ref{qualitative_parc_results}. The conventional procedure of aligning different brain surfaces to a reference is an expensive computational step. Our method learns this alignment step automatically by capturing the geometry of the brain, yielding a Dice overlap of 83.2$\%$. Qualitatively, as illustrated in Fig. \ref{qualitative_parc_results}, the performance of our method is similar to a GCN trained with traditional alignment, however computation times are reduced by a 1400-fold, from 15 seconds to 10.7 \textit{milliseconds}. The use of SGT is evaluated in this paper with brain surface parcellation as an application. Nevertheless, our method can potentially be used for other surface analysis problems such as disease classification or identifying new geometry-related biomarkers. 
 
\medskip
\noindent
\textbf{Acknowledgment} -- This work was supported financially by the MITACS Globalink Internship Program, the Fonds de Recherche du Quebec (FQRNT), the Research Council of Canada (NSERC) and NVIDIA with the donation of a GPU. 

%
\addtolength{\textheight}{-6cm}
\bibliographystyle{IEEEbib}
\bibliography{Reference}


\end{document}